# Designing a 9-channel location sound microphone from scratch

Florian Camerer     (*flocamerer@gmail.com*)

**Abstract**

The design of a 9-channel microphone system for location recording of mainly atmospheres will be described. The key concept is matching the recording and reproduction angles of the individual sectors. The rig is designed for the AURO-3D 9-channel playback system (4 height speakers). An analysis of the reproduction layout will be included, as well as recording concepts like the Stereo Recording Angle (SRA), Williams curves, Scale Factors for different reproduction angles than 60° and diffuse field decorrelation. Finally, practical aspects like microphone mounts and windshields for such a system will be presented.

## 1. Introduction

Recording sound on location is mostly understood as capturing speech/dialogue in case of feature films and Television and sound effects as well as backgrounds to be used as elements in sound design. Other applications such as soundscapes bring the capturing of a whole sound scene, of the atmosphere of a place more into focus. The different applications feature specific esthetics as well as reproduction environments. This has a direct impact on the microphone arrangements used. For movies, recording backgrounds or atmospheres has almost exclusively been done in Mono and 2-channel Stereo, especially before the widespread use of discrete Surround Sound (5.1, 6.1, 7.1). The situation in Television is quite similar. Soundscape recordings have sometimes utilized Soundfield microphones to capture all three dimensions ("Ambisonics"), but it is only recently that greater interest can be acknowledged in microphone systems which can deliver original recordings for Immersive Audio. In this paper, we will specifically look at designing such a microphone system, covering basic localization parameters, diffuse field decorrelation and the actual mechanical realization.

## 2. Constraints, framework, choices

- We will design the system for the AURO-3D 9-channel reproduction layout [1]. With its 4 loudspeakers for the height channels directly above the corresponding loudspeakers of the basic 5.1 layout (ITU-R BS.775-2 [2]) it provides a reasonable extension of the most common planar Surround Sound setup. Nevertheless, a few details regarding the layout need to be addressed. They especially concern the reproduction angle of the Height loudspeakers (see chapter 2.1).

- We will try to optimize the following sound engineering parameters:
    - Envelopment
    - Stability (large sweet spot); detachment from the speakers
    - Localization of direct sources, horizontally and vertically

- The microphone system should be handled by a single person. The following practical parameters are relevant:
    - Size, weight, portability
    - Easy setup ("plug & play")

- o  Flexibility in positioning the microphones
- o  Decoupling from handling and wind noise

Let us look at each parameter in more detail.

## *2.1    The AURO-3D reproduction layout*

AURO-3D 9.0 is a recording and reproduction system for Immersive Audio, based on the standard 5.1 ITU-R BS.775-2 layout, adding 4 loudspeakers for the height information. These are often positioned directly above the respective ear-level Base layer speakers as shown in **Figure 1**:

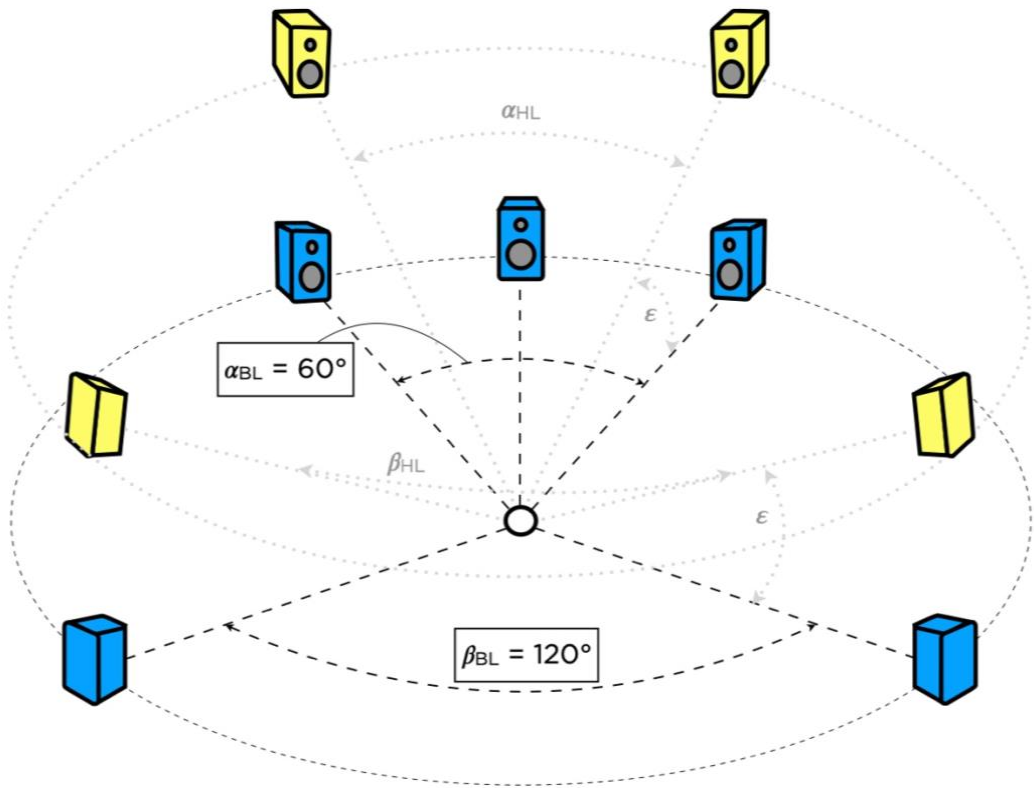

**Figure 1**: AURO-3D 9-channel loudspeaker layout

For the Base layer we will choose the angle between the Surround loudspeakers to be 120° which is at the limit of the tolerance described in ITU-R BS.775-2. With this choice the result is an equilateral triangle between the Center and the Surround speakers. If the Height loudspeakers are positioned directly above the respective Base layer speakers, the enclosed angle between pairs of speakers is reduced, depending on the elevation angle of the Height speakers ($\varepsilon$ in Fig.1). Straight forward trigonometry gives us the enclosed angle for adjacent pairs of Height loudspeakers:

$$\alpha_{HL} = 2 \arcsin [\sin(\alpha_{BL}/2) \cos \varepsilon] \qquad \text{[Eq. 1]}$$

$\alpha_{HL}$ ….. Angle between the listening position and an adjacent Height loudspeaker pair

$\alpha_{BL}$ ….. Angle between the listening position and the respective Base layer loudspeaker pair

$\varepsilon$ ……... Elevation angle of the Height speakers

For $\alpha_{BL}$ = 60° (Base Front Left and Right speaker) and $\varepsilon$ = 30°, the enclosed angle of the Height Front speakers $\alpha_{HL}$ equates to 51.3° which is significantly narrower than the Base angle of 60°. Similarly, for a Base layer Surround speaker angle of $\beta_{BL}$ = 120°, the respective angle of the Height layer Surround loudspeakers $\beta_{HL}$ becomes 97.2°, more than 20° narrower than the Base layer reproduction angle!

If we want to have the same reproduction angles between at least the two Base Front and the two Height Front loudspeakers, we have to modify the positioning of the Height speakers (a complete matching of the reproduction angle of all pairs of adjacent Height speakers to their Base counterpart pair is not possible, as the sum of all four angles for the elevated loudspeakers with the listening position is smaller than 360°). For the Height Front loudspeakers, let us first assume that the speakers are positioned in the same plane as the Base Front speakers. Any two speakers mirrored at the vertical plane through the listening position and spanning 60° lie on two cones with their tips at the listening position and facing directly left and right (opening angle 120°). These two cones intersect with the plane through the Front loudspeakers in the form of a hyperbola ($h_1$ in Fig.2). Every point in space being 30° elevated from the listening position lies on a cone facing directly upwards with an opening angle of 120°. Also this cone intersects with the plane through the Front loudspeakers with a hyperbola ($h_2$). Where the two hyperbolas intersect the two new positions of the Height Front speakers can be found (see **Figure 2**).

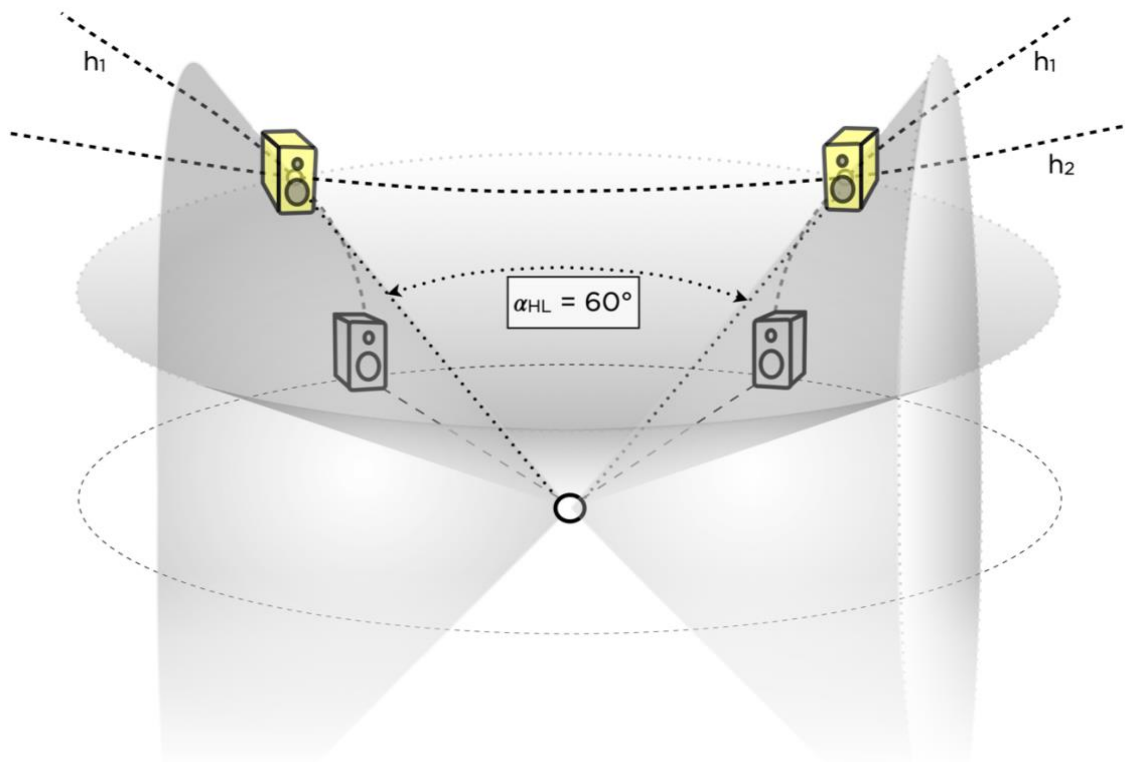

**Figure 2**: Compensation of the position of the Height uniplanar Front speakers to obtain a 60° reproduction angle

Also the elevation of the Height speakers is consequently adapted slightly compared to the position directly above the Base layer speakers. In the non-compensated position the elevation of the Height speaker (for an elevation $\varepsilon$ of 30°) is the radius r of the layout multiplied by $\tan\varepsilon$ which is approximately 0.58r. For the compensated Height speaker position this becomes 0.61r (the increased distance from the listening position is approximately 1.22r, which can again be calculated with trigonometry).

*Example:*
Base layer speaker angle $\alpha_{BL}$ = 60°; elevation $\varepsilon$ = 30°; speaker layout radius = 3 m
Elevation of compensated Height Front loudspeaker = 1.84 m (measured from the level of the Base layer speaker). This is 0.11 m higher than the uncompensated position;
Increased distance from sweet spot = 3.66 m (0.33 m more to the side in the plane of the speakers). The time compensation needed is 1.8 ms (0.00064r).

For the Height Surround speakers, we find that there is no practical solution if we position the Height loudspeakers in the same plane as the ones from the Base layer and if we want a 120° reproduction angle also for the Height Surround pair. The two hyperbolas (similarly to the Front situation described above) intersect in infinity. Therefore we accept the compromise of positioning the Height Surround speakers directly above the Base layer speakers (which often is a practical necessity anyway as the Surround speakers are placed next to the wall due to constraints of the size of the listening room). This leads to a deviation in our general approach of matching the Height and Base layer reproduction angles (at least for the Front and Surround speaker pairs), as the Height Surround angle is now 97.2° (described above; time-compensation = 0.00044r or 1.3 ms for r = 3 m)). The design of the microphone system takes this deviation into account.

A different layout approach is based on a hemispherical setup, keeping the distance of all Height loudspeakers the same from the listening position (no time-compensation needed). When we also keep the reproduction angle the same as the respective Base layer speaker pair, the Height speakers theoretically move on a circle perpendicular to the plane of the Front or Surround speakers and intersecting with the speakers (similar to the side-facing cones in Fig.2). This again works only satisfactorily for the Height layer Front speakers as only for these there exist distinctive intersecting points of those circles with the "elevation cone" in Figure 2. For the Height layer Surround speakers, the respective circles touch the elevation cone at 90° to the side of the listening position. Speakers in this position are arguably not "rear" Surround speakers. Also a reasonably coherent sound field with the Base Surround speakers is more difficult. Again, accepting a narrower reproduction angle for the Height Surround speakers allows more flexibility.
In practical terms, this approach necessitates separate speaker stands for all four Height loudspeakers. For the compensated layout described above, at least the Height Surround speakers are using the same stands as the Base layer speakers. Therefore we decide to choose this compromise: position-compensated Height Front speakers, Height Surround speakers directly above the Base layer speakers, time compensation for all four Height speakers.

## 2.2 Envelopment

Studies have shown that the perception of envelopment is linked to the amount of decorrelation especially in the low-frequency range [3] [4] [5]. There is a direct dependency of the lower limit of decorrelation of two microphone signals on the distance and/or polar pattern of the microphones. For coincident 2-channel Stereo systems, only supercardioids, hypercardioids and figure-of-eight arrangements can provide a Diffuse Field Correlation Coefficient of zero [6] (see **Figure 3**). For a spaced pair, decorrelation is achievable with any polar pattern. Wittek's *Image Assistant* (see link in chapter 3.1) provides a comprehensive illustration of the different results depending on the distance. The more a microphone exhibits the characteristics of a *pressure* transducer (100% being an omnidirectional pressure microphone), the better the low frequency performance is (the theoretical frequency response of a pure pressure transducer can go down to 0 Hz). A pressure *gradient* transducer like the figure-of-eight has practical limitations regarding its frequency response due to the decreasing pressure gradient towards lower frequencies.

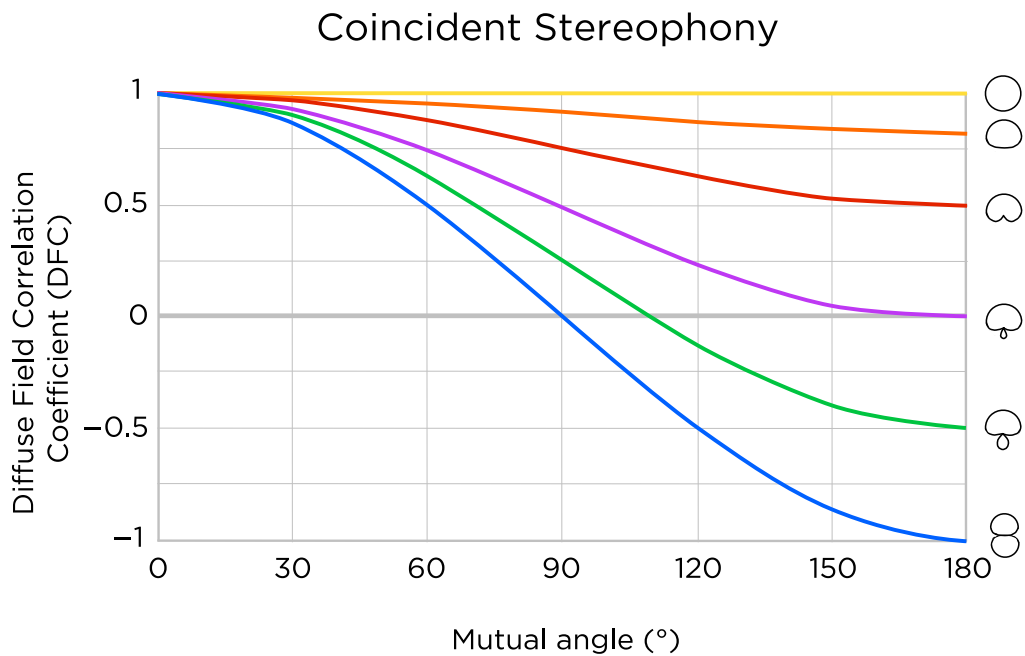

**Figure 3**: Diffuse Field Correlation Coefficient vs. microphone angle for Coincident Stereophony

We will choose the **wide cardioid** (or Hypocardioid) as our polar pattern for all nine microphones. It has 2/3 omni characteristics (thus exhibiting good low frequency performance) and still a reasonable directivity with 10 dB attenuation at 180°. The wide cardioid also exhibits a quite constant polar pattern up to at least 8 kHz (for small diaphragm microphones). Furthermore, we choose an arrangement with **spaced microphones**, as in a coincident system with wide cardioids we cannot obtain a Diffuse Field Correlation Coefficient of zero.

## 2.3 Stability, *sweet spot size; detachment from the speakers*

Unfortunately, there are two mutually exclusive goals for any Stereophonic microphone system regarding its result over loudspeakers considering direct sources:

- o Large sweet spot, stable localization over a wide listening area
- o Reproduced sound "detaches" from the speakers, phantom images provide the majority of the auditory experience

In the first case, the sounds in every speaker need to be highly decorrelated, making the actual loudspeaker quite noticeable. This is the result of a large distance between the microphones (often omnidirectional), leading to a narrow Stereo Recording Angle (SRA, see chapter 3.1) and only minor phantom imaging.

In the second case, the Stereo Recording Angle is usually deliberately chosen to match the spread of the source in front of the microphones, so that the loudspeaker base is filled with phantom images, making the speakers "disappear" perceptually. For this effect to happen, the listener needs to be placed exactly equidistant between the (two) loudspeakers, and the size of this sweet spot is quite small. This need not be a disadvantage, if one is "living in the sweet spot" like Michael Williams, the main protagonist of deliberate Stereo Recording Angle choice [7]. Nevertheless, for our 9-channel location microphone system, we will opt for a compromise between the two extremes, providing good decorrelation for diffuse sounds, a reasonably sized sweet spot as well as noticeable phantom imaging with good detachment from the loudspeakers the closer one sits to the center of the reproduction system.

## 2.4  Localization of direct sources

The translation of a 2-channel Stereophonic microphone system to a loudspeaker system has been studied for some time (see, for example, [8] [9]). Although there is no consensus in the scientific community regarding the exact values of Inter-Channel Level Differences (ICLDs) and Inter-Channel Time Differences (ICTD) needed for covering the whole reproduction angle (for example, 60° in the ubiquitous equilateral Stereo-triangle), we will use this information to determine the angles and distances for the microphones of our Immersive Audio rig.

We will design our system based on the following main concept: *the Stereo Recording Angle of any two horizontally adjacent microphones shall be the same as the reproduction angle of the respective loudspeaker pair which plays the signals of those two microphones.*

This theoretically results in a 1:1 relationship of the source position and its corresponding phantom image. In reality, phantom imaging is not equally good around a listener in a reproduction layout like 5.0 Surround Sound or AURO-3D 9.0. It is best in the front, a bit worse in the back (due to the large angle between the Surround loudspeakers), much worse to the sides (one ear cannot provide a phantom image) and basically non-existent in the vertical plane (see below). For moving sources (like, for example, birds, insects, cars, people walking by etc.), the localization difficulties are not as severe, especially if the movement is fast.

In the vertical plane, the mechanisms of ICLD and ICTD work "differently" regarding phantom imaging between any two vertically spaced loudspeakers [10]. Localisation is often governed by the pitch-height effect and thus frequency-dependent [11] [12]. For ICLDs, Lee has done studies giving guidance for the amount of level difference needed between the Height and Base layer of loudspeakers in an Immersive Audio reproduction setup to achieve separation between the layers regarding direct sounds. For 0 ms ICTD, the suggested minimum level difference is 9.5 dB, for ICTDs in the range of 1-10 ms, the minimum level difference is 7 dB. As we have chosen wide cardioids, this theoretically necessitates a vertical angle of 135° between the Height and Base layer microphones to achieve 7dB of level difference for sources hitting the main axis of any one microphone. We will reduce this angle to **90°** because we also position the Height microphones vertically spaced, thus providing additional level differences for close sources. This situation can often be found in outdoor environments (for example, insects on a meadow flying close by).

## 2.5 Practical considerations

- *Size, weight, portability:* As we want to design our system to be handled by one person, this apparently limits the physical attributes of the rig. While this is highly subjective, it sets useful constraints on the dimensions and material as well as equipment choices. Our real-world system (see chapter 3.5) is arguably close to the upper extreme of the scale (the total weight including tripod and multichannel recorder is 20 kg)…

- *Easy setup ("plug-&-play"):* We will design our rig with a single multicore cable and a corresponding break-out box attached to the recorder. The windshield construction is attached to a base plate which snaps into a tripod head. The setup should be finished within 2 minutes. Attention is directed also to the carrying solution (backpack frame), making the transition from transportation to ready-for-recording as straight forward and fast as possible.

- *Flexibility in positioning the microphones:* Although we will design our system with one single arrangement in mind, we still want the flexibility to adjust the position of the microphones, in case we want to record for a different loudspeaker layout, or in case new findings make finetuning of microphone positions feasible. The microphone mounting thus has to allow this flexibility. The same holds true for the distance between the Height and Base layer.

- *Decoupling from handling and wind noise*: without good microphone mounts and an effective and acoustically transparent windshield outdoor location recording is compromised or impossible (for example, when recording strong wind). Although individual windshields for each microphone are a viable solution (and frankly, in most cases the only one), we will use a single windshield for each layer. This diminishes the effect of, for example, a wind gust hitting mostly one windshield and thus revealing the respective loudspeaker. Such a single windshield per layer needs to have a certain size, apparently. We will use a custom-made solution for this purpose.

## 3. Realization of the 3D location audio microphone setup

### 3.1 Localization, Stereo Recording Angle

The main concept we will use for the actual dimensions of the microphone layout is the *Stereo Recording Angle (SRA)*. This can be understood as being the invisible angle in front of any 2-channel Stereophonic microphone arrangement that is mapped to the reproduction angle of the respective loudspeaker pair. If a source spans the SRA, its phantom images will completely fill the Stereo base. In music recordings, usually the SRA is adapted to the width of the source. Here we adapt the SRA to match the reproduction angle. The position and/or movement of a source is subsequently translated faithfully on the reproduction side.
To calculate the SRA of a pair of microphones, a few tools are available, for example the *Williams' Curves* (http://mmad.info/MAD/2%20Ch/2ch.htm), the *Image Assistant* by Wittek (http://ima.schoeps.de/), the *Microphone Array Recording and Reproduction Simulator (MARRS)* iOS and Android App by Lee et al. and the *MARRS for the Web* by Goddard and Lee (http://marrsweb.hud.ac.uk). These tools greatly help the design of microphone systems as they intuitively provide the trade-off relationships of ICLD and ICTD. A disadvantage is that these tools all provide slightly different prediction results due to different values for the image shift corresponding to an ICLD of 1 dB or an ICTD of 0.1 ms. In practical terms, this is not dramatic, as the results are close enough.

We will use *Image Assistant* to begin our design. Starting with the Front 3-channel Base layer microphone arrangement, we get the result shown in **Figure 4** for an SRA of 2x30° using three wide cardioids, an angle of 60° between the side facing microphones and the main axis and an offset of 14 cm of the Center microphone (our deliberate choice).

The horizontal span of the three front microphones is calculated as 200 cm. This is arguably quite large for a location rig that should be portable. More troublesome is the necessarily even larger windshield which we want to contain all five Base layer microphones. With additional space to provide effective wind noise suppression, such a windshield would be at least 210 cm wide. This would be clearly impractical and also mechanically unstable.

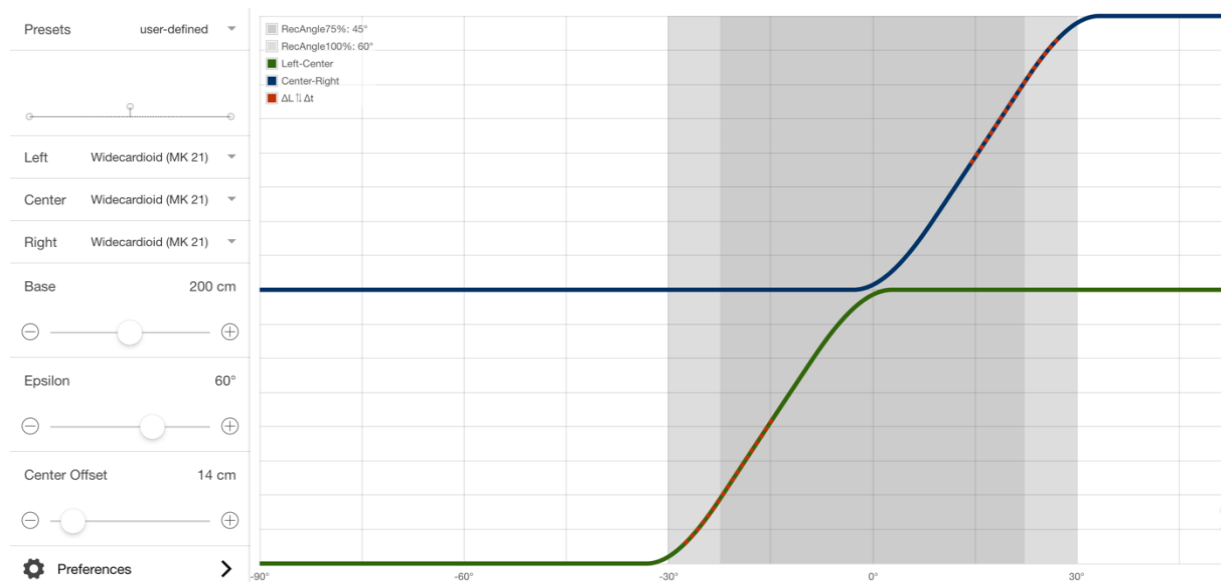

**Figure 4**: Dimensional results of an SRA of 2x30° with three wide cardioids, using *Image Assistant*

Fortunately, there is a solution that does not enforce us to compromise.
This solution is based on the findings of Lee and Rumsey [13] where the Stereo base is divided into two shift regions: the first spans 0 – 66.7%, the second spans 66.7 – 100%. The tradeoff in shift factors for the first region is 13.3°/0.1 ms for ICTD and 7.8°/1 dB for ICLD. For the second region, the shift factors are 6.7°/0.1 ms and 3.9°/1 dB, half of the values for the first region. This differentiation is due to the uncertainty of panning towards the outer limits of the reproduction angle. While this concept is not intrinsically different than the solutions of Williams and Wittek, it has been extended by Lee [14] introducing *Scale Factors* for reproduction angles different from 60° (*Williams' Curves* and the *Image Assistant* map their calculation results on any loudspeaker pair's base angle). **Figure 5** shows the proposed linear ICTD and ICLD trade-off functions by Lee et al. [14] based on the data from Lee and Rumsey [13]:

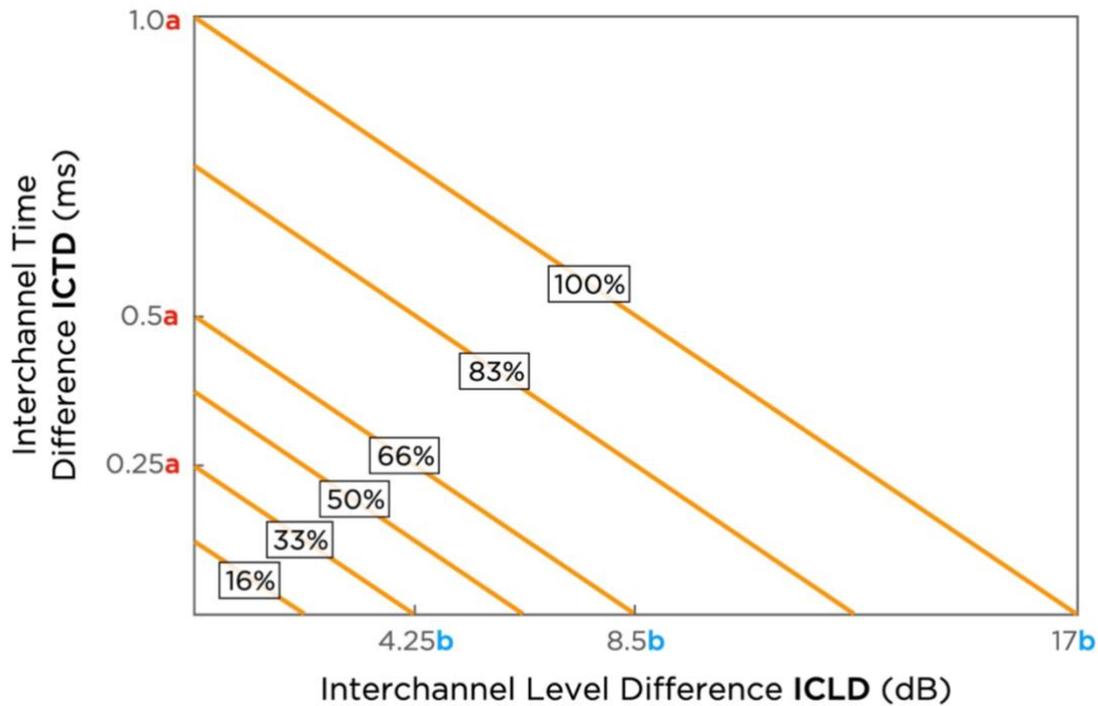

**Figure 5**: Proposed linear ICTD and ICLD trade-off functions
(according to Lee and Rumsey); see later for Scale Factors **a** and **b**

The Scale Factors for reproduction angles (loudspeaker base angles) different from 60° are determined by the Interaural Level Difference (ILD) and the Interaural Time Difference (ITD) at 60° and the ILD and ITD of the different speaker base angle. For the AURO-3D 9.0 reproduction layout used for our design, the Base layer Front angle is 2x30°, the Surround angle is 120° and the Side angle is 90°. We are therefore interested in the Scale Factors for 30°, 90° and 120° for the Base layer.

For the Height layer, our compensated Front angle is 60°. This does not necessitate any scaling and allows the use of all the aforementioned tools (although MARRS for the web will provide the Scale Factors integrated in the app). The omission of Scale Factors is one of the reasons for the compensated positioning of the Height Front loudspeakers. The Height Surround angle is approximately 97.2°, the Side angle is about 75.5° (using Eq.1 with $\alpha_{BL}$ = 90°).

Generally, applying the Scale Factors for the ICTD-ICLD trade-off functions for speaker base angles different from 60° matches well with our listening experience. **Figures 6** and **7** show the different Scale Factors **a** and **b** for the angles of the Base and Height layer (based on equations in [14] and graphs in [15]):

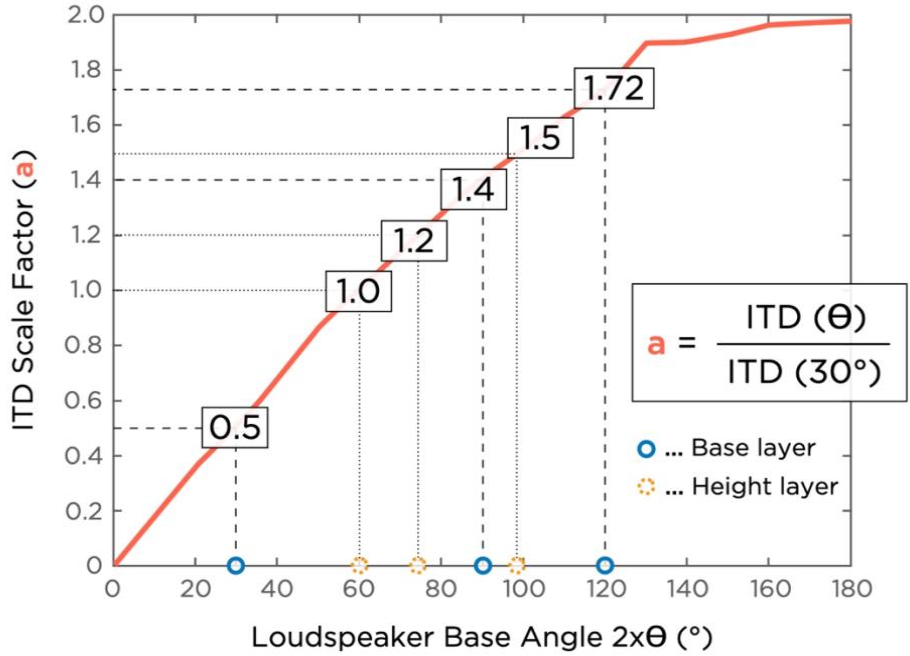

**Figure 6**: ITD (Interaural Time Difference) Scale Factor **a** for arbitrary loudspeaker base angles (based on [14], [15]); blue solid circles on the x-axis indicate the angles of the AURO-3D Base layer speaker pairs, orange dashed circles the angles of the Height layer speaker pairs

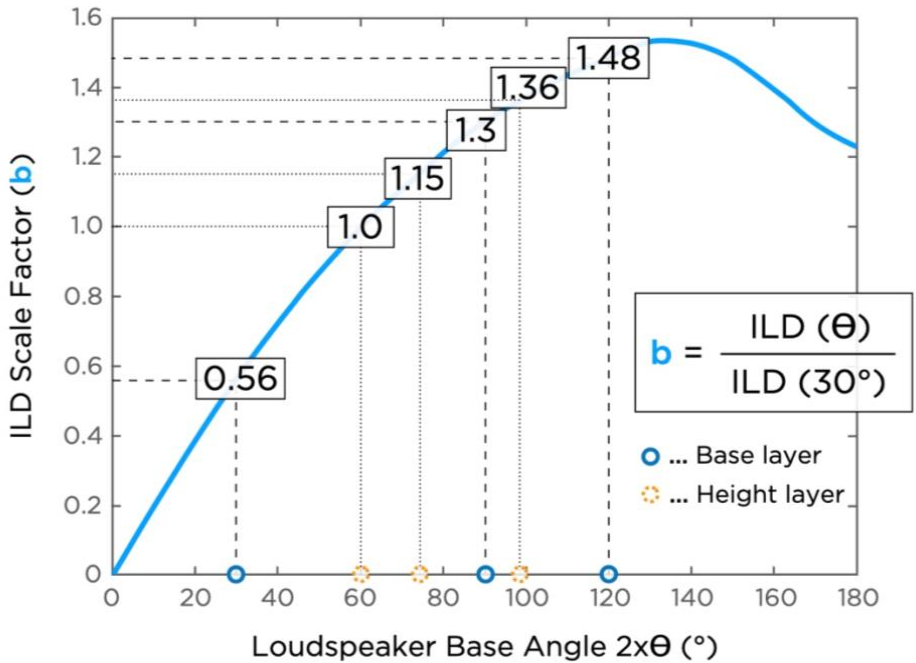

**Figure 7**: ILD (Interaural Level Difference) Scale Factor **b** for arbitrary loudspeaker base angles (based on [14], [15]); blue solid circles on the x-axis indicate the angles of the AURO-3D Base layer speaker pairs, orange dashed circles the angles of the Height layer speaker pairs

With our choice of using wide cardioids as the polar pattern for the microphones and spacing them apart to achieve decorrelation down to low frequencies, the main localization mechanism is based on ICTD. *Image Assistant* and *MARRS* show the relationship between ICTDs and ICLDs (see **Figure 8** for a screenshot of the relevant page of *Image Assistant*). Consequently, weighting the ICTD contribution by about 2/3 we take a combined factor of 0.5 (Scale Factors for 30°) for the Base Front 3-channel microphone layout[1]. This results in a reduction of the distance between these three microphones to **100 cm**, considerably more manageable from a practical viewpoint than 200 cm.

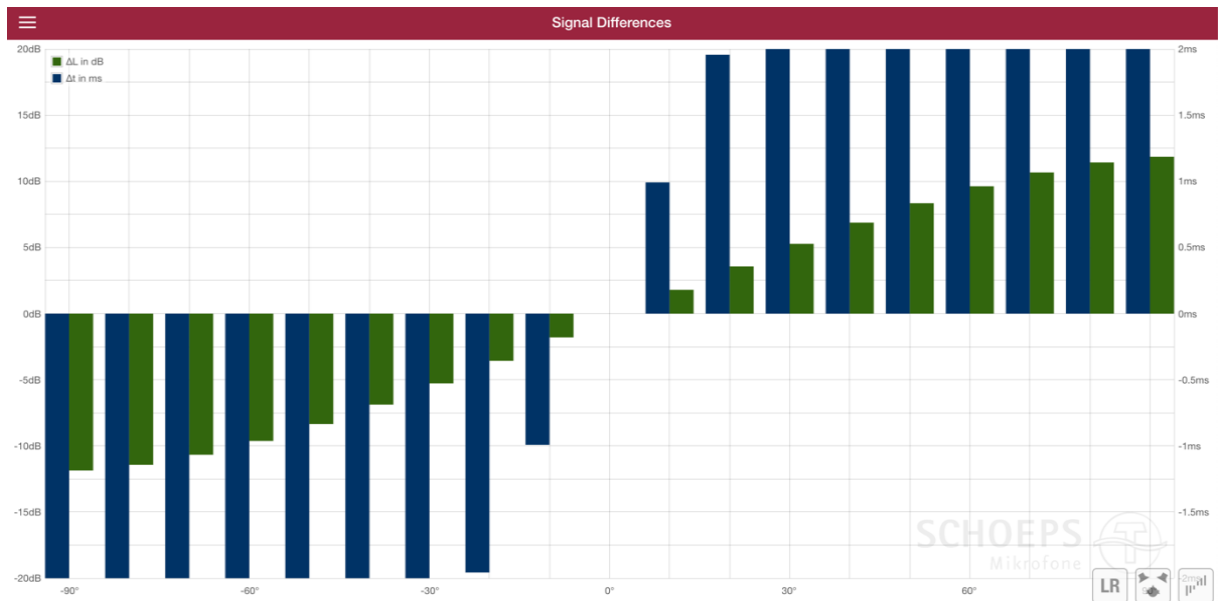

**Figure 8**: ICTD and ICLD for three wide cardioids as shown by *Image Assistant*
(200 cm span, 60° angle of left & right microphone, 14 cm offset of center microphone)

For the Surround Base layer microphone pair (120° SRA and reproduction angle) we derive a Scale Factor of (almost) 1.7 which increases the distance of the two Surround microphones for the Base layer from 24 cm (44° offset from the main axis; original dimensions derived with *Image Assistant*) to approximately **41 cm**.

The side SRA and reproduction angle of 90° gives a necessary scaled distance between the two wide cardioids of about **48 cm** (the angle between a Front side-facing microphone and the adjacent Surround microphone is 76°). This lets us calculate the distance of the Surround microphones from the axis of the left and right Front microphone. Using Pythagoras' theorem, we get about 39 cm for this distance (or 53 cm from the center microphone to the axis between the left and right Surround mic). The SRA is tilted to the back and does not seamlessly touch the Front and Surround SRAs (something Williams calls "critical linking"- he solves this using level and/or time offsets for individual microphones). As localization to the sides is unstable anyway and due to practical constraints, we accept the compromise of applying no compensation.

**Figure 9** shows the final Base layer microphone arrangement:

---

[1] This "weighting" is a hypothesis of the author. Lee has agreed to perform calculations if such a weighted combination of the ICTD and ICLD Scale factors is adequate.

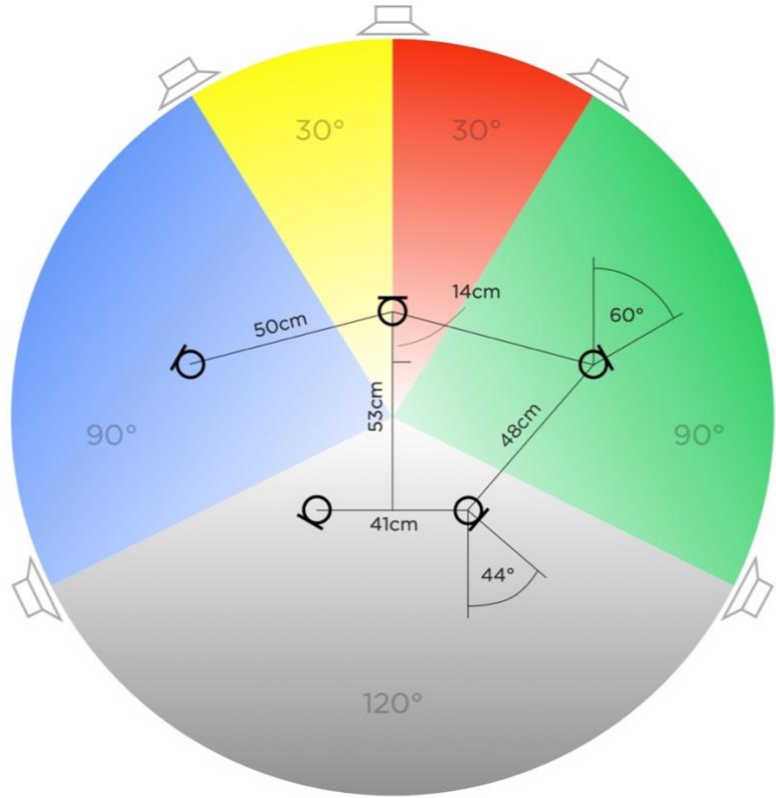

**Figure 9**: Dimensional details of the Base layer microphone arrangement for an AURO-3D 9.0 reproduction layout; the reproduction angles and Stereo Recording Angles of respective pairs of microphones and speakers match

We conduct a similar process for the Height layer microphones (also wide cardioids). The lack of a Height Front Center speaker is advantageous from a size standpoint, as the Height recording and reproduction angle is 60° (compensated layout) and not 2x30°. This decreases the dimensions of the Height layer significantly and does not need any Scale Factor for the Height Front pair. We choose a distance of **50 cm** and an angle of 2x43° (many combinations are possible for a 60° recording angle). The Height Surround pair is calculated with the relevant Scale Factor (1.45) for 97.2° and a choice of an angle between the microphones of 2x44°. The resulting distance is about **43 cm** (30 cm without Scale Factor, obtained with *Image Assistant*). The side pair with its recording and reproduction angle of about 75° and the enclosed angle of 93° as a result of the choices for the Front and Surround pair gives us a compensated distance of about **46 cm**. **Figure 10** shows the final Height layer microphone arrangement. The indicated distance of the two lines connecting the Height Front and Height Surround microphones respectively is also about 46 cm (a little bit less, actually) due to the similar spacing.

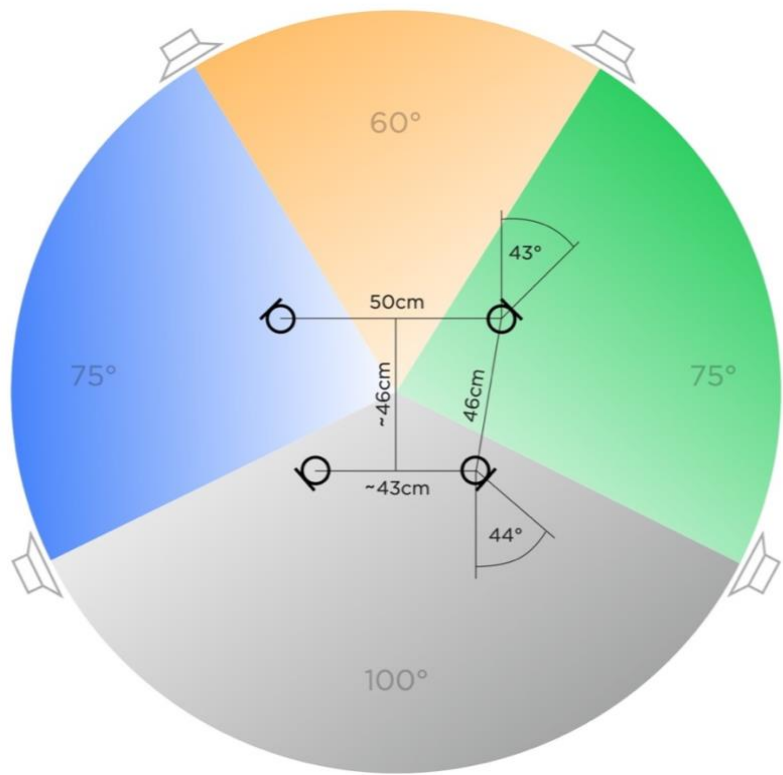

**Figure 10**: Dimensional details of the Height layer microphone arrangement for an AURO-3D 9.0 reproduction layout; the reproduction angles and Stereo Recording Angles of respective pairs of microphones and speakers match; due to the elevation of the loudspeakers, the sum of all four angles is < 360°, in this case about 310°

## 3.2   Diffuse Field Correlation Coefficient (DFC)

While we would like a DFC of zero down to the lowest frequencies for optimized envelopment, we have to accept a compromise because of the dimensions of the arrangement. For example, two adjacent (Base layer) Front microphones (50 cm apart) provide a DFC of zero starting from slightly above 300 Hz upwards. For the left and right microphones (almost twice the distance), the lower frequency limit becomes about 150 Hz. At 75 Hz, we still have a DFC of 0.5. See **Figure 11** for the case of the Left and Right (Base layer) Front microphones (here the parameter Diffuse Field Image predictor (DFI) is used which is similar to the DFC; DFI is defined in [16]).

A larger arrangement would provide even lower decorrelation for the diffuse field but would lead to narrower recording angles and thus more focus to the speaker positions. This concept can be used for applications where comprehensive phantom imaging is not important, such as for feature films. Here, larger dimensions result in an almost completely decorrelated diffuse field with enhanced stability in larger listening rooms (such as cinemas).

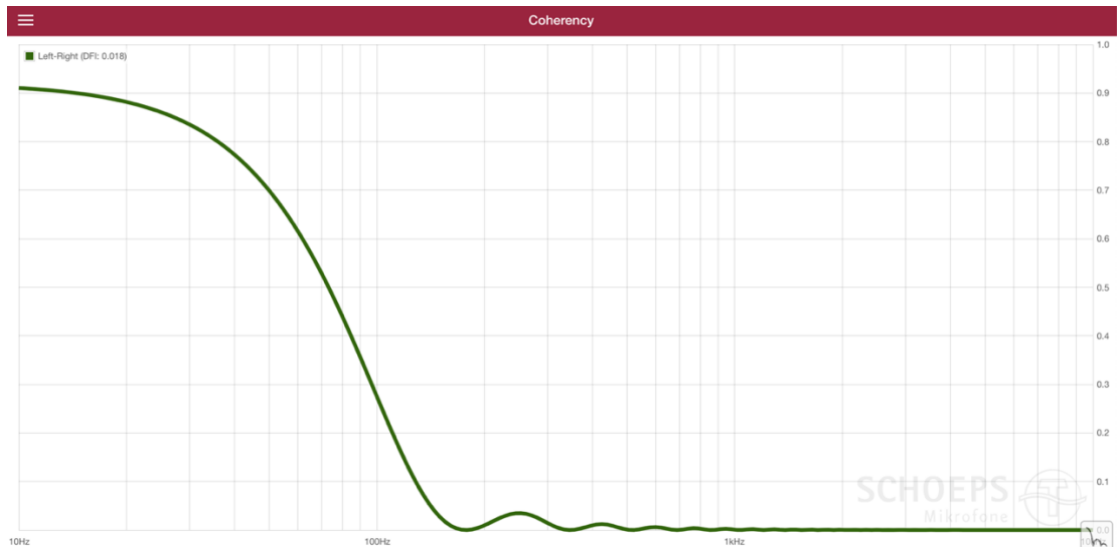

**Figure 11**: Diffuse Field Image predictor (DFI) for two wide cardioids 100 cm apart

### *3.3    Distance Height - Base layer microphones*

In chapter 2.4 we described our choice of spacing the Base and Height layer microphones to increase the level differences for close sources. Another reason for the vertical spacing is described by Gribben and Lee [17]. In a study about the perception of band-limited decorrelation between vertically oriented loudspeakers they found that vertical decorrelation is only effective above about 500 Hz. When we look at the Diffuse Field Image predictor of two wide cardioids angled 90° (which is our choice for the angle between the Base and Height layer microphones), we can see that decorrelation for the frequency range above 500 Hz becomes almost zero for a spacing of 30 cm (calculated with *Image Assistant,* similarly to Figure 11). Consequently, this spacing is the maximum needed vertical distance. If due to practical reasons a slightly lower or higher spacing seems appropriate, it will shift the "decorrelation cut-off frequency" (as Gribben & Lee call it) a bit. In the case of a higher cut-off frequency this can even be an advantage to avoid lower frequency spectral distortion and a compromise in tonal quality. Therefore we choose a distance **25-30 cm** between the Base and Height layer microphones.

### *3.4    Summary of dimensions*

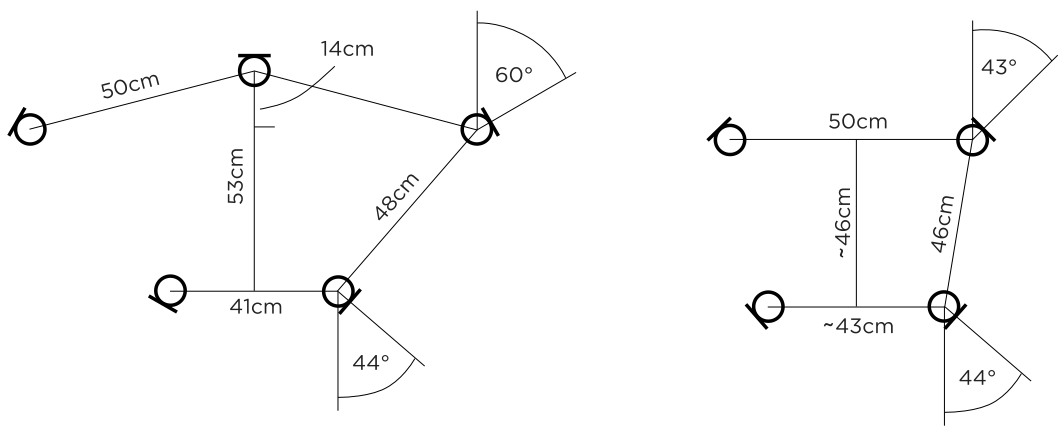

9 wide cardioids   /   spacing Base-Height layer: 25-30 cm   /   Base-Height mic angle: 90°

## 3.5 Structural realization

Our choice of two large single windshields makes a custom solution necessary. We are grateful that Philippe Chevenez from CINELA (http://www.cinela.fr/catalog.php) has designed and built such a solution. As one windshield alone arguably looks like a UFO, the complete 3D-rig has been nicknamed "**Double-UFIX**" (**Figure 12**). With its additional fur cover the structure provides good suppression of wind rumble, and the elastic mounts decouple the microphones down to 50 Hz from handling noise.
The Double-UFIX is mounted on a two-layer custom base plate out of plywood and aluminum which rigidly connects to a camera tripod (including fluid head) for stable positioning and flexible orientation.
The 9 microphone signals are connected to two small boxes inside the windshields and then fed into a single multicore which terminates at a breakout box with individual connectors for the multichannel location recorder. The multicore solution facilitates a quick setup.

The whole rig can be carried on a backpack frame. A connecting plate like on the head of the camera tripod is mounted on the frame, so that the Double-UFIX with its base plate attached can be easily fixed on the frame. For transportation, the tripod is slid into a dedicated tube with a lid, also attached to the frame (**Figure 13**). The system allows a setup time < 2 min as indicated in chapter 2.5. **Figure 14** shows the system during an atmosphere recording session on location.

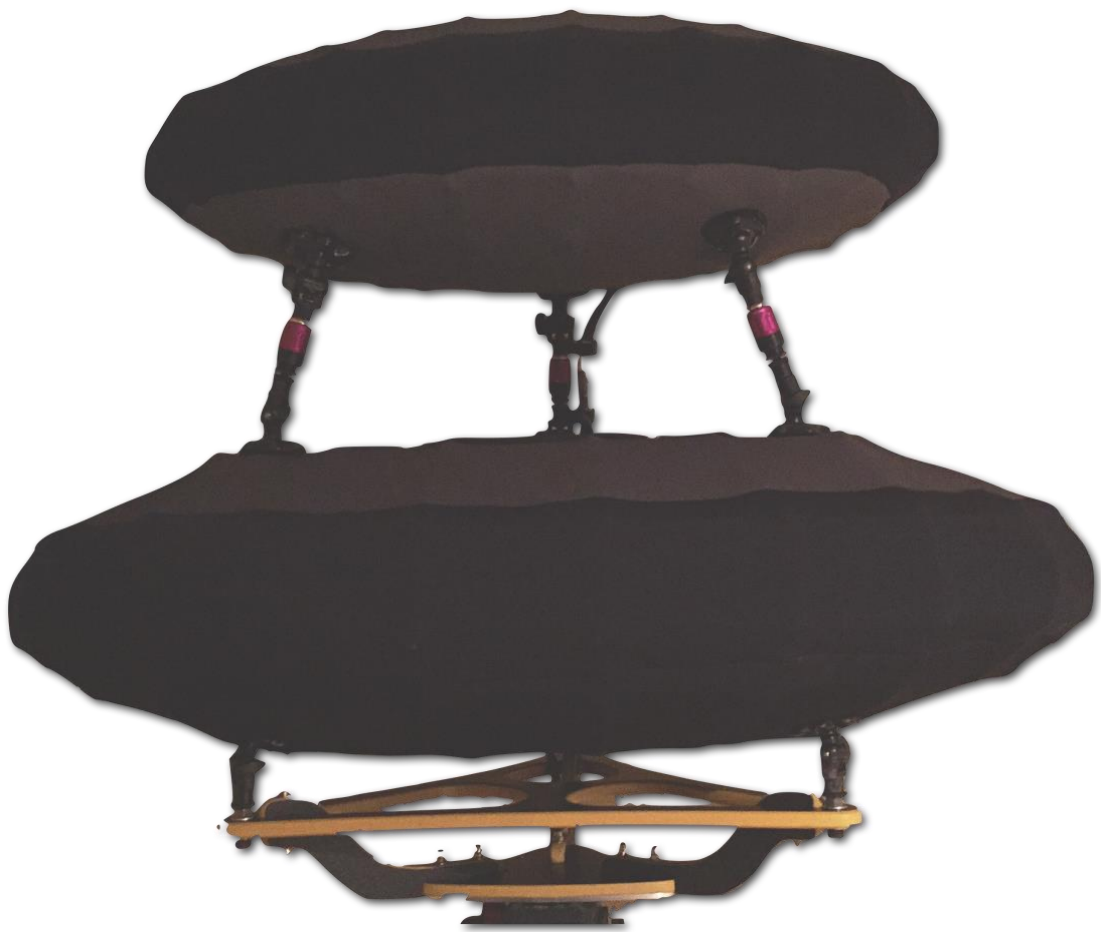

**Figure 12**: Double-UFIX windshield with base plate

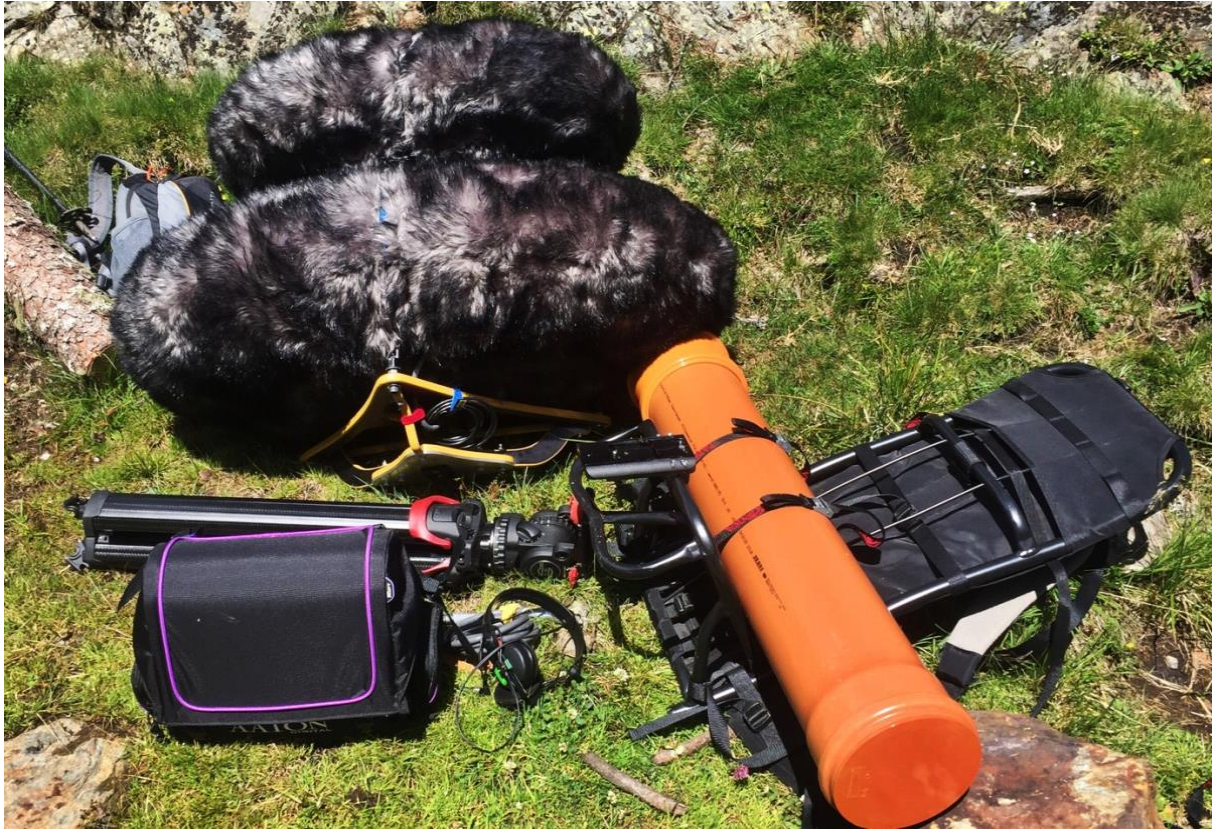

**Figure 13**: Double-UFIX rig with tripod, backpack frame and recorder

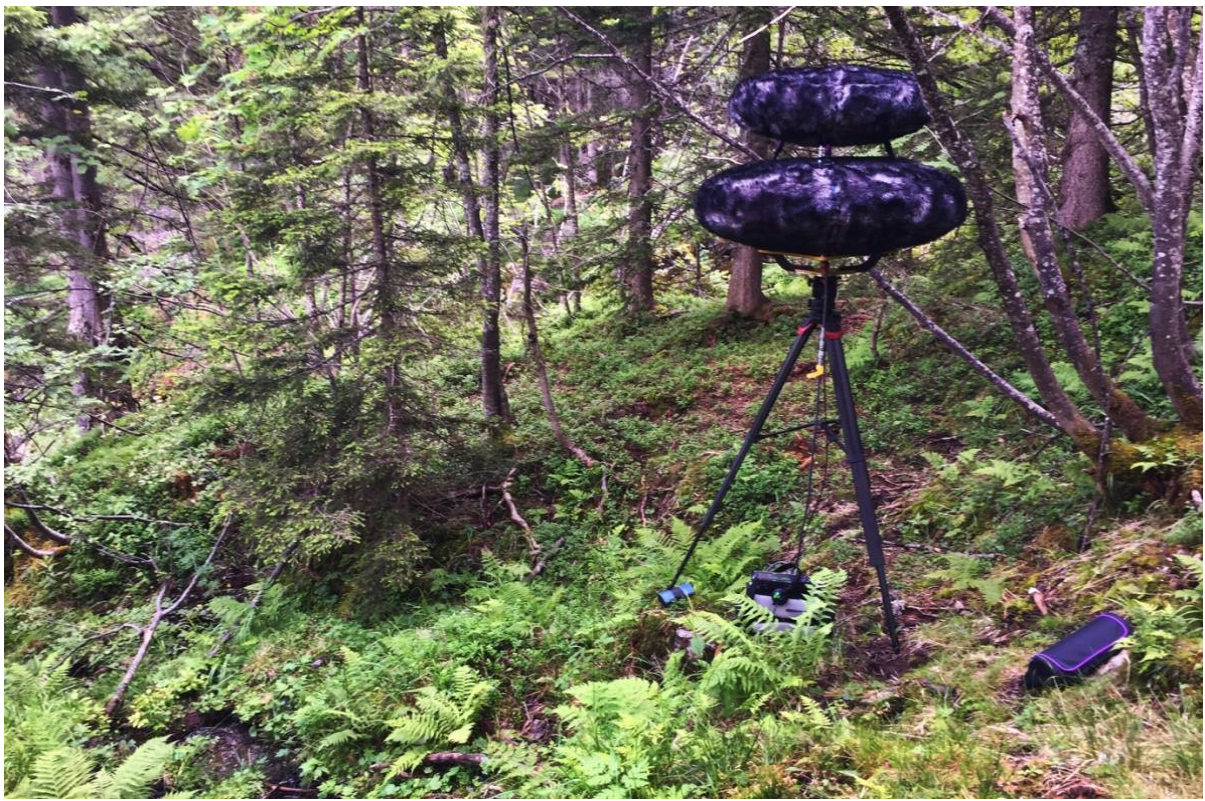

**Figure 14**: Double-UFIX system on location

## 4. Summary

We have designed a 9-channel Immersive Audio microphone arrangement for mainly location recording of atmospheres. The approach was to match the Stereo Recordings Angle of any two horizontally adjacent microphones with the reproduction angle of the respective loudspeaker pair. First practical experiences show that this delivers a convincing illusion of the auditory environment. The goal was to minimize the angle distortion of the source positioning during reproduction as well as providing as much decorrelation as possible for the diffuse field. This fully applies to the horizontal plane; in the vertical plane, a spacing of 25-30 cm is used to achieve decorrelation above about 500 Hz (below that frequency, decorrelation is not effective in the vertical plane).

We have also described the structural realization of such a microphone system. It is arguably close to the limit of what a single person can carry and handle. For easier transportation, the next step will be to attach wheels to the carrying frame, so that the rig can be pulled.